# Mechanical-microstructural correlation on SPS-fabricated NiTi alloy

Tadeáš Těhan[1], Jaromír Kopeček[2], Elizaveta Iaparova[2], Eduardo Alarcón[2] and Sneha Samal[2, *]

[1]Faculty of Electrical Engineering, Czech Technical University, Prague
[2]Institute of Physics of the Czech Academy of Sciences, Na Slovance 2, 182 00 Prague, Czech Republic

**Abstract**

Mechanical properties and dynamical mechanical analysis were performed on compact Spark Plasma Sinter samples. It has been observed that the sample with less porosity reflects the behavior of superelasticity response. Other samples show failure during first cycles that may be due to porosity. Compaction of metallic powder is one of the standard procedures in the field of powder metallurgy for the fabrication of bulk material. Consolidation of the sample as a function of sintering temperature plays a crucial role in the final compaction mechanism. However, the evolution of compaction, microstructure, phase transformation and mechanical properties as a function of sintering temperature is hardly disclosed. In this work, a correlation has been established between mechanical and microstructural properties of compact samples. The maximum compactness and the corresponding microstructure, porosity, texture, phase transformation, grain size, hardness, mechanical properties of compact samples were discussed. The research establishes mechanical properties-structure correlations for compaction of NiTi alloy in advanced engineering applications.

Corresponding Author: Sneha Samal, Email:samal@fzu.cz

## 1. Introduction

The conventional applications of NiTi are shifting from standard biomedical to advanced engineering fields. There is an increasing trend to fabricate NiTi samples using powder metallurgy to use various technologies rather than purchasing the material from a standard supplier. In the powder metallurgy sector, tuning the sample size, geometries, and shape is a more open challenge based on the demand of the application. Spark Plasma Sintering (SPS) is one of the standard, simple, common technologies that allow for compacting powder into bulk form using both temperature and pressure. The quality of the compact fabricated materials estimates the sintering efficiency of the SPS process. The efficiency depends on the sintering parameters of the process, i.e., temperature, pressure, current, voltage, holding time, ramp rate for heating and cooling. The most crucial parameter is that the vacuum chamber facilitates a fully dense compact material during the sintering process. The parameters also influence the sintered product microstructure and mechanical behavior. Our earlier articles [1-2] found that densification was mainly influenced by the particle size and temperature. The sintering temperature plays a significant role in the formation of a compact product. By increasing the temperature, the porosity reduces, and the samples become more compact. On increasing the sintering temperature, the heat travels from the surface to the core, which results in the softening of the boundary of particles to form a larger contact surface. The particle size and sintering temperature play a major role in Cu powder densification by the SPS technique [3]. The intrinsic driving force, local pressure, and current of powders also depend on the particle size, which inherently leads to densification in the sintering process. The relative density decreases with an increase in particle size and also influences the sintering mechanism during the process [4]. Increasing the sintering temperature, the density was increased for various materials such as NiTi alloy, Ti alloy, and W alloy processed by SPS techniques [5-6]. Thus, the compactness and microstructure that lead to mechanical properties are significantly affected by the sintering temperature.

Higher densification products can be achieved through the SPS process by tuning the parameters of sintering technologies. The SPS method is simple, faster, and easy to manipulate the technological aspects to control the heating profile [7-8]. During SPS, heat is generated by the flow of pulsed current and voltage through the graphite punch, maintaining the vacuum environment. The generation of plasma causes a cleansing action by promoting the connecting particles on the softening grain boundary. However, the generation of plasma in the SPS method is still an open question [9-10]. SPS is considered one of the remarkable methods for synthesizing and compacting both transitional and novel materials [11]. Earlier researchers used a sintered product, that on increasing sintering temperature at fixed load, the samples became more and more compact with a decrease in porosity [12-14]. When two particles contact with each other, temperature plays a significant role in necking by softening, melting and evaporation mechanisms [15-19].

In this work, the compaction of NiTi alloy has been investigated and a correlation has been established between the sintering mechanism of densification with microstructure evolution. The microstructural evolution such as porosity, grain size distribution, phase and element composition of the compact product is systematically investigated and discussed. The texture evolution has been studied on the miniature samples cut from the bulk sample in transverse and rolling directions. The phase transformation behavior, mechanical properties were investigated in the tensile behavior, at room temperature. The storage modulus, loss modulus, tan delta were measured and dilatation behavior of the fabricated samples were tested during thermal cycles in 3-point bending mode.

## 2. Experimental Methods Characterization

### 2.1. Fabrication of NiTi samples

The powders of NiTi (50 at. %, 99.5 % purity) were purchased from American Elements (AE), USA, and were subjected to compaction by the SPS technique. The alloy samples were fabricated at a load of 50 MPa at various temperatures of 900, 1050, 1100 and 1150 °C by SPS, DR. SINTER™ 1030 machine, 2.7 ms DC pulse (1–3 V, 1000–2000 A). The as-fabricated sample underwent grinding and polishing by standard metallographic procedure to remove the graphite contamination from the die and chamber.

### 2.2. Characterization of fabricated samples

The microstructures of powder and samples were characterized using a FEI Quanta 3D Dual-Beam SEM/FIB from Thermo Fisher Scientific (Brno, Czech Republic). The grain size of the sintered sample was observed with a Keyence optical microscope after grinding, polishing, and etching with Kroll's reagent (5 mL $HNO_3$, 10 mL HF, and 85 mL $H_2O$). Grain orientation and phase distribution were analyzed using electron backscatter diffraction (EBSD) with an EDAX DigiView V camera. The cross-sectional surface of coating layers on the substrate of mild steel was prepared by cutting the coating with the substrate using an electric discharge machine. Samples were hot-mounted in conductive bakelite and polished with colloidal silica before finishing with Kroll's reagent for grain size measurements. The Image J software was used to measure the particle size of NiTi powder and annealed powder. The miniature tensile samples were prepared by laser cutting from the bulk samples in both transverse and radial directions. The corresponding texture evolution has been studied as a function of grain size and two different directions of transverse and rolling in compact samples. The mechanical properties of compact NiTi alloy were investigated in tensile mode in a dynamic mechanical analyzer using stress control operation. The miniature samples were prepared by laser cutting from bulk samples in two different orientations, such as the transverse direction (TD) and rolling direction (RD). The samples were tested in Dynamic Mechanical Analysis (DMA) in tensile mode to understand the mechanical behavior. The corresponding orientation of grains in TD and RD samples, with investigation of texture evolution, was estimated at various degrees of crystallographic orientation along <111>, <100>, and <110>. Simultaneously, two samples from TD and RD of sample 1150 were chosen to investigate the grain boundary maps and disorientation, and harmonic texture at various orientation directions in the NiTi alloy. Microhardness was measured using a Vickers hardness tester at various loads (245.2 mN) for 10 s per measurement [15]. The austenite, martensite, and R-phase transformation temperatures were investigated using Differential Scanning Calorimetry (DSC, 25, TA Instruments, New Castle, DE, USA) at 5 K/min from −100 to +100 °C, using samples of approximately 15-23.5 mg in aluminum pans. Uniaxial tensile tests were performed using an Instron ElectroPuls 10,000 testing machine. All tests were conducted under displacement control at a constant crosshead speed of 0.01 $mm \cdot s^{-1}$. The specimens had a gauge length of 10 mm. Axial strain was measured using a virtual extensometer based on real-time digital image correlation (DIC). The virtual extensometer had a gauge length of 5 mm and was used to continuously monitor the axial deformation throughout the tests. Mercury RT software was used for real-time DIC processing. Before mechanical loading, the specimens were heated to 100 °C by resistive (Joule) heating through the application of an electric current while maintaining a target load of 0 N to minimize thermally induced stresses. After reaching 100 °C, the specimens were allowed to cool naturally to room temperature before tensile testing. This thermal pre-treatment was applied to ensure a well-defined and reproducible initial microstructural state for all specimens before mechanical loading. The most compact sample was chosen for bending analysis by a thermo-mechanical analyzer and dynamical mechanical analyzer. Thermomechanical loading tests were performed using a DMA-850 tester (TA Instruments) in tensile mode. Rectangular plate specimens with a gauge length of 10 mm and a cross-section of 0.65*0.35 mm were loaded at 180 °C to a constant tensile stress of 10 MPa. Subsequently, the samples were cooled from 180 °C to −140 °C at a rate of 10 °C/min while subjected to a sinusoidal oscillation at a frequency of 10 Hz. After reaching 140 °C, the specimens were heated back to 180 °C under the same loading conditions. The experiments were carried out in force and temperature control mode to investigate the thermomechanical response of the material during thermal cycling. The superimposed dynamic oscillation enabled continuous measurement of the dynamic mechanical response throughout the thermal cycle, allowing changes in the elastic properties in the material associated with the temperature-induced phase transformation to be monitored. The dilation tests were performed on the most compact samples SPS 1100, SPS 1150 in both fabricated and annealed samples under 3-point bending mode at 200 mN load under thermal cycles at the rate of 5 K/min under thermal cycles.

## 3. Results and Discussion

### 3.1. Pre-alloy powder NiTi

Fig. 1 displays the SEM image of NiTi ( 50 at. %) and annealed NiTi powder morphology. The powder size distribution is represented by the histogram of particle size as the function of relative frequency and cumulative frequency. The phase transformation temperatures were presented from the thermal cycles as the function of cooling and heating cycles. The powders are spherical in shape and size with an average particle diameter of $d_{avg}$: 43.10 ±0.75 µm and $d_{90}$: 52.17 µm. However, the annealed powders (annealed at 500 ºC, 1 h) show a slight agglomeration of particle size $d_{avg}$: 43.18 µm and $d_{90}$: 52.8 µm.

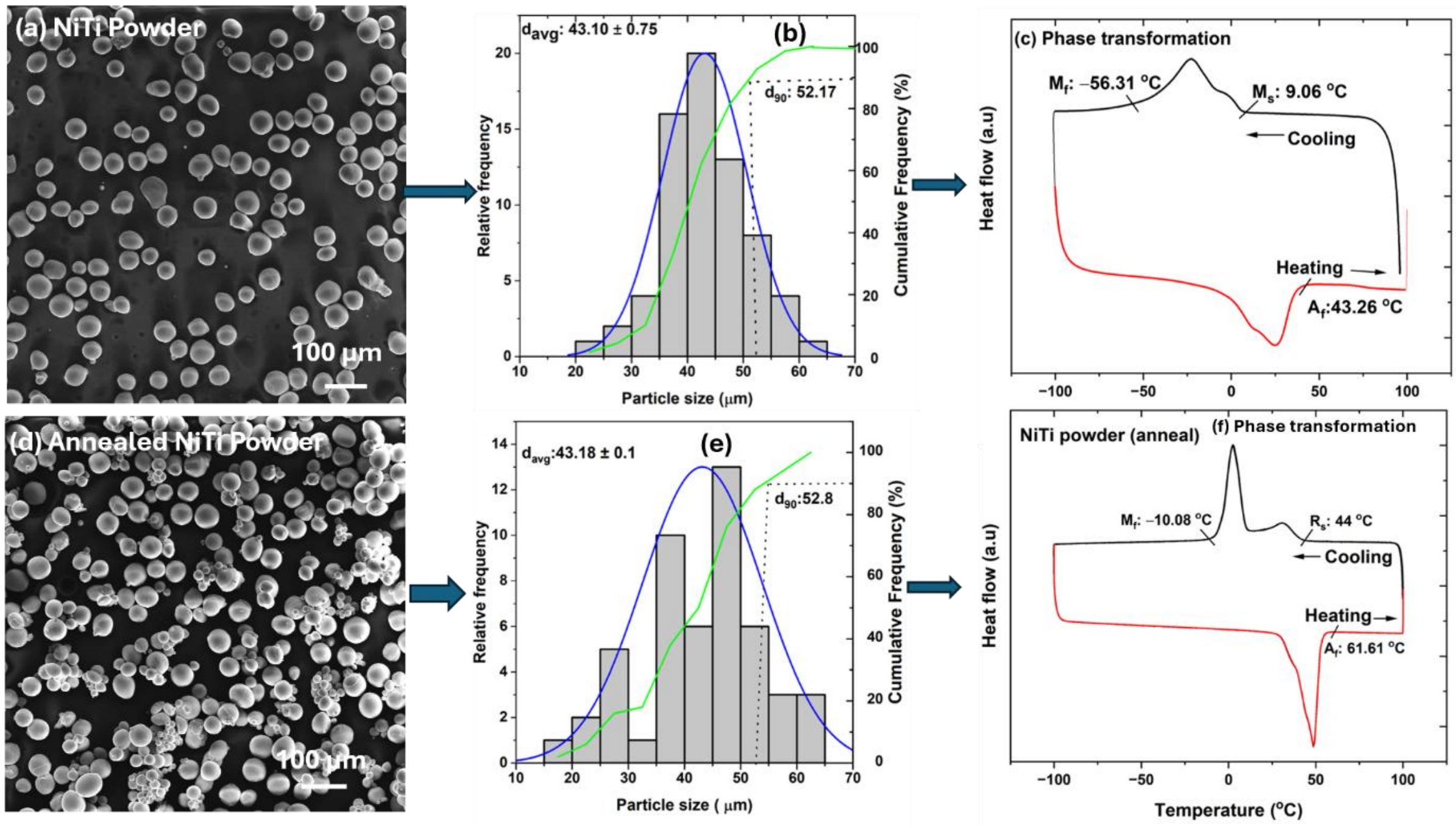

Fig. 1. NiTi powder characteristics: a) Size and shape of the NiTi particles characterized by SEM, b) Histogram of powder size distribution ($d_{avg}$: 43.10 µm, $d_{90}$: 52.17 µm), c) DSC scan of the NiTi powder performed to evaluate transformation temperatures ($M_s$ =9.06 ºC, $A_f$=43.26 ºC) d) Size and shape of the annealed NiTi powders ( 500 ºC , 1 h) characterized by SEM, e). Histogram of annealed powder size distribution ($d_{avg}$: 43.18 µm, $d_{90}$: 52.8 µm), f). DSC scan of the NiTi powder performed to evaluate transformation temperatures ($R_s$ = 44 °C, $A_f$ = 61.61 °C).

### 3.2. Fabrication of NiTi alloy by SPS

Fig. 2 a) portrays the experimental profile of the sample preparation and b) shows the schematic diagram of the SPS system with a graphite crucible that carries NiTi powders with a die and a punch. There is a load that is allowed to be applied in contact with the die. The whole assembly was placed inside the SPS chamber to undergo the sintering process at various sintering temperatures at fixed load. The chamber was depressurized to maintain the vacuum and initiate the process. The electric pulsed current and voltage were initiated to achieve the desired temperature. The profile displays the ramping stage to achieve the sintering temperature for a fixed period of time, 10 min at a constant pressure of 50 MPa. The chamber was depressurized by releasing the load and allowed to cool down in the ambient environment. The sample was released from the graphite chamber and underwent grinding to remove the graphite. Then the samples were prepared by laser cutting in both the transverse direction (TD) and radial direction (RD) using a power of 7 W for 10 min. Table 1 represents the sample specifications and geometries considered for this study. The sintering parameters and laser cutting parameters are listed for both samples. The samples are very similar in thickness, ranging from 0.6 to 0.7 mm. The miniature samples were prepared from the bulk sample by laser cutting in both transverse and radial directions to the normal axis. NiTi powders are placed between the graphite dies. The whole assembly was transferred to SPS chamber. The load was applied and kept fixed, followed by programming the sintering profile.

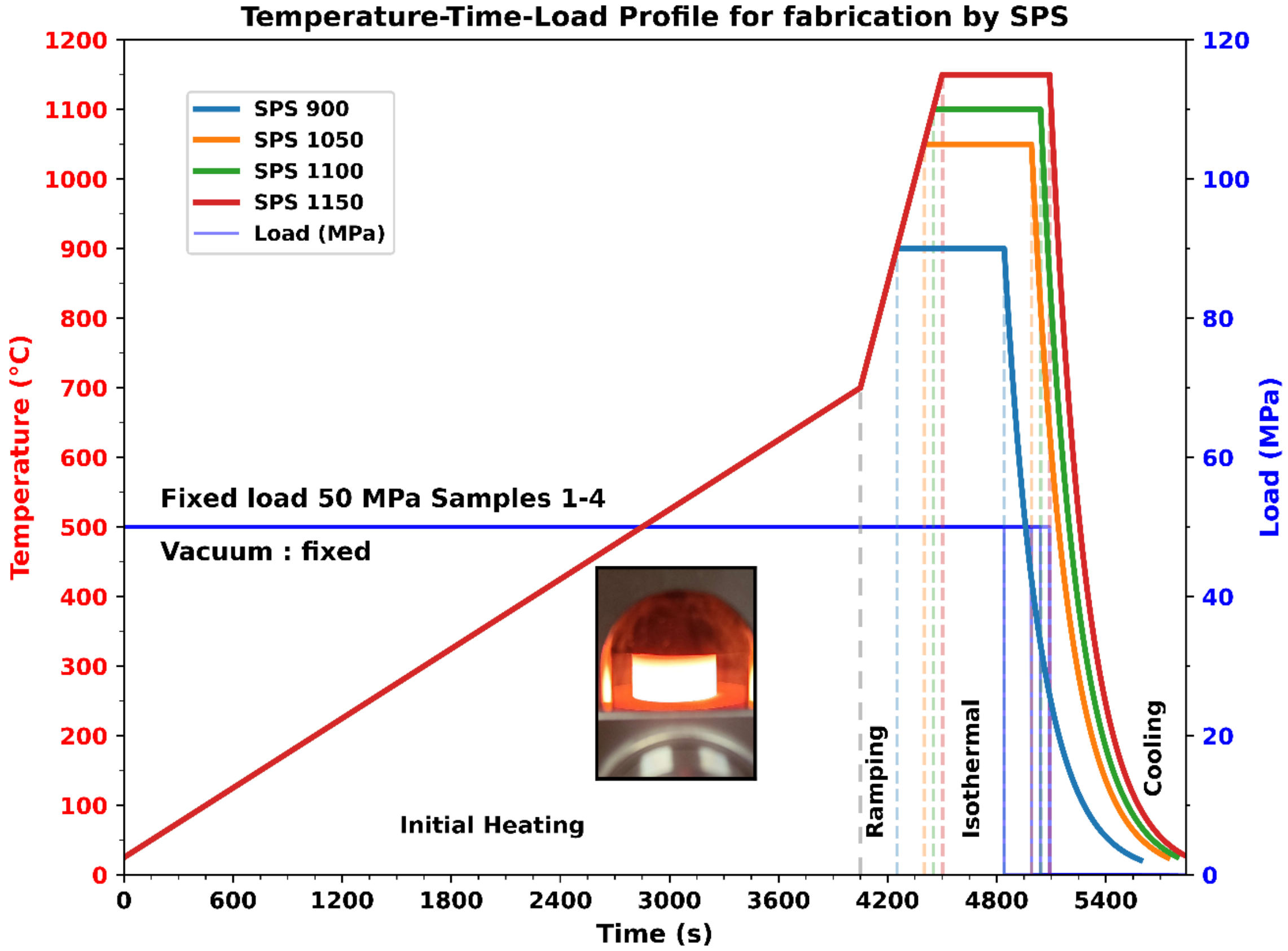

Fig. 2. Temperature-time-load profile of sample fabrication by SPS system at 4 different sintering temperatures 900, 1050, 1100 and 1150 ºC at fixed load of 50 MPa ( with inset image of active sintering process), inset image of SPS chamber.

**Table 1** represents the SPS parameters for the preparation of samples.

| Parameters | Sample 900 | Sample 1050 | Sample 1100 | Sample 1150 |
|---|---|---|---|---|
| Sintering temperature | 900 °C | 1050 °C | 1100 °C | 1150 °C |
| Thickness | 0.6 mm | 0.65 mm | 0.68 mm | 0.7 mm |
| Compaction | 94.96 % | 96.47 % | 99.21 % | 99.80 % |
| Load | 50 MPa | | | |
| Powder for compaction | Spherical, 5-55 µm, 99.9 % purity | | | |
| Dog bone sample for symmetry observation | Sample 4, Laser power to cut the sample: 7 W, Spot size: 100 ns, Repetition: 600 s (TD, RD) | | | |

### 3.3. Microstructural grain size distribution and compositional analysis in SPS samples

The microstructural analysis of the SPS samples was investigated along the cross-section to observe the compactness and the voids. It has been observed that samples sintered at 1150 °C show no voids and are denser and more compact compared to other samples. Fig. 3 (a-d) displays the microstructural image of samples 900-1150 in secondary electron mode. It has been observed that triangular-shaped voids appear in sample SPS 900, which increase the quantity in SPS 1050, which are marked red in color. However, sample SPS 1100 shows pinhole-sized voids in the sample that develop at the joining of two adjacent grain boundaries. The triangular voids ( marked in red) generated in sample SPS 900 vanish in sample SPS 1150 upon increasing sintering temperature. These voids, considered as necking regions among grain boundaries, diminish in more compact regions, which is observed in sample SPS 1150. The grain size distribution in samples was revealed in Fig. 4 ( a-d) with more pronounced phases of austenite and martensite regions. The grain diameter is distributed in the range of 5-55 µm in all the samples. The average grain size of sample SPS 900 is more scattered from the range of 5-55 µm, similar to NiTi powder. However, sample SPS 1100 has an average grain size of 15 µm, and sample 1150 shows smaller grains with an average grain diameter of 10-12 µm. The sintered sample exhibits a polycrystalline grain size distribution of smaller and some coarse grains. The grain sizes are reduced in sample SPS 1150, which may correspond to a higher sintering temperature, which may allow the particles to undergo better compaction with a wider neck region to form coalescence. The grain size distribution along the cross-section areas and the corresponding phases of the austenite and martensite region are investigated and presented in Fig. 4.

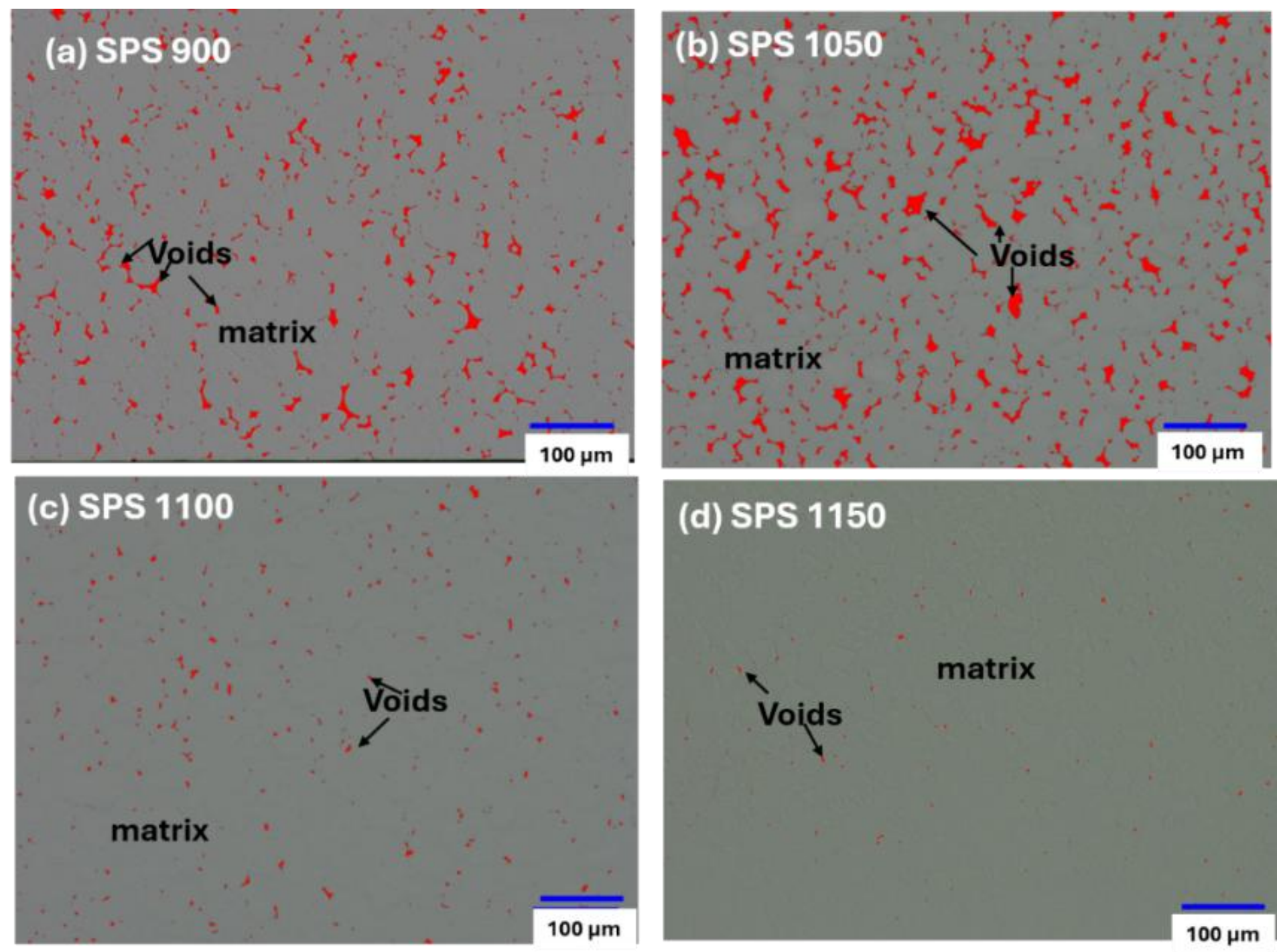

Fig. 3. Cross-sectional image analysis of the fabricated sample shows a) SPS 900 ( sample fabricated at 900 ºC) shows the matrix in gray region (grains) and voids are marked in red color b) SPS 1050 ( sample fabricated at 1050 ºC) shows the matrix gray region (grains) and voids are marked in red color, c) SPS 1100 ( sample fabricated at 1100 ºC) shows the matrix gray region (grains) and voids are marked in red color, d) SPS 1150 ( sample fabricated at 1150 ºC) shows the matrix gray region (grains) and voids are marked in red color.

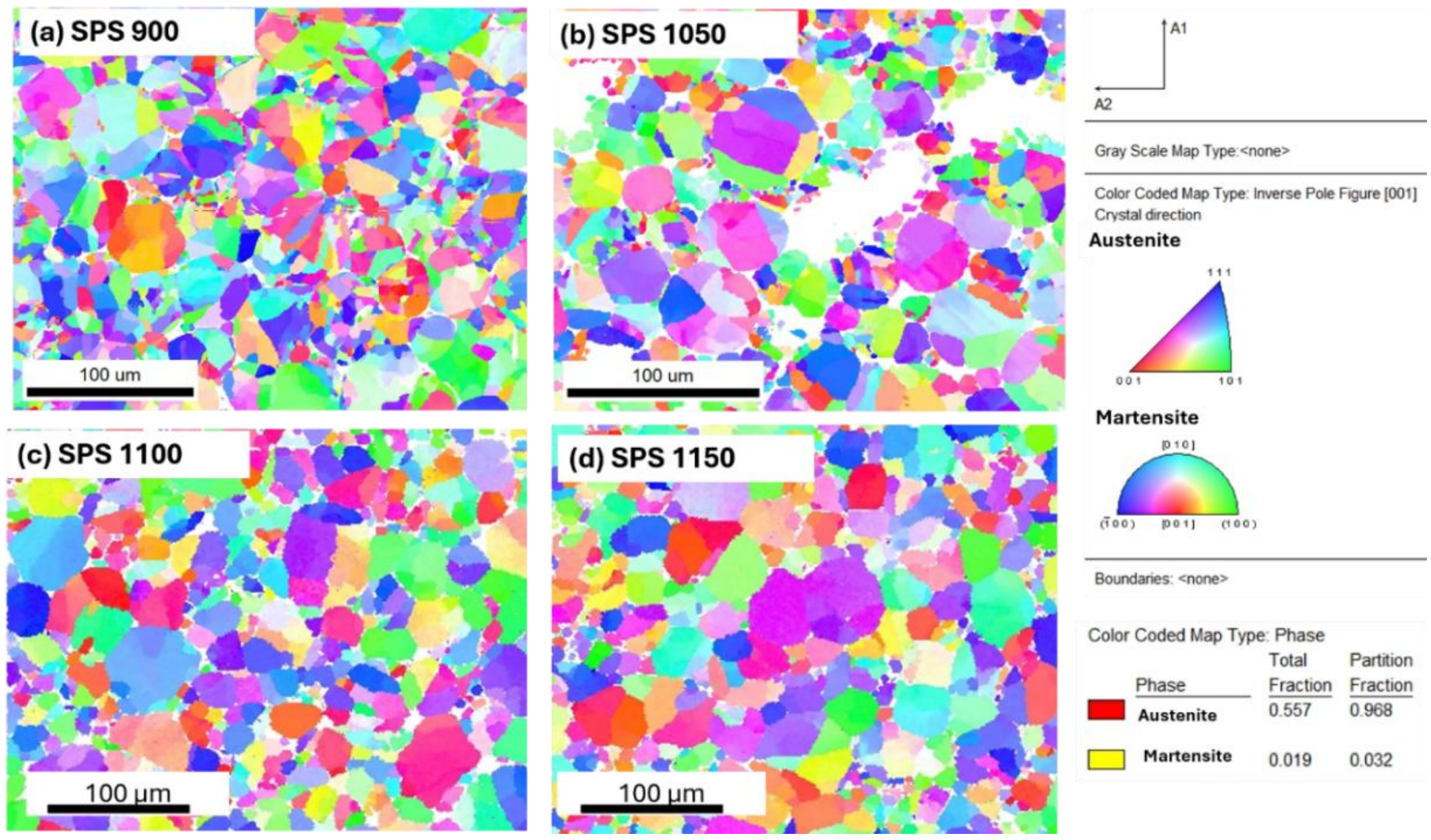

Fig. 4. Grain size distribution and the corresponding phases of austenite and martensite are shown in a ) SPS 900; b) SPS 1050; c) SPS 1100; d) SPS 1150. The inverse pole figure shows the region of austenite (96 %) and martensite (3.2 %).

The findings of the phase fraction by EBSD analysis match well with XRD quantitative analysis. The samples are austenite dominant phases of 96 Wt % and martensite is 3.2 % in fraction areas. As the SPS 900 and SPS 1050 samples show a large number of voids, to investigate the compositional analysis, we chose the most compact samples, SPS 1100 and SPS 1150, for EDX analysis. Table 2 represents the chemical composition analysis of the most compact samples. The chemical composition of the sample shows that SPS 1100 has Ti: 47.9 Wt % and Ni: 52.1 Wt. % and SPS 1150 shows an amount of Ti: 47.8 Wt. % and Ni: 52.2 Wt. %. Both samples have very similar chemical composition, with a Ni-rich alloy.

Table 2: Chemical composition of sample SPS 1100, SPS 1150 by EDX analysis.

| Sample | Element | Weight % | Atomic % | Net Int. | Error % | R | A | F |
|---|---|---|---|---|---|---|---|---|
| SPS 1100 | Ti | 47.9 | 53.0 | 7158.6 | 1.9 | 0.5736 | 0.9687 | 1.0497 |
| | Ni | 52.1 | 47.0 | 3737.6 | 2.2 | 0.6410 | 0.9767 | 1.0364 |
| SPS 1150 | Ti | 47.8 | 52.9 | 7105.3 | 1.9 | 0.5736 | 0.9687 | 1.0498 |
| | Ni | 52.2 | 47.1 | 3722.2 | 2.2 | 0.6410 | 0.9768 | 1.0364 |

### 3.4. Porosity, grain size, compaction and hardness of the samples

The quantitative distribution of porosity, grain size distribution, compactness, and hardness of the fabricated samples were evaluated and presented in Fig. 5. Fig. 5 (a) represents the SPS 1050 has maximum porosity compared to other fabricated alloys, which may be due to the smoothening of grain boundaries that leads to necking and increasing contact areas between the grains. As the neighboring two grains coalesce with each other, the porosity enlarges and gradually decreases with an increase in sintering temperature. The porosity reduces and is minimized in the SPS 1150 sample, which is more compact in nature. Fig. 5 (b) displays the grain size distribution in the sintered fabricated samples by SPS. The grain sizes are in the range of 5-55 μm, similar to NiTi powder for the fabricated samples of SPS 900 and 1050. However, the compact samples of SPS 1100 and 1150 contain more distribution of smaller grains that may develop during compaction and higher sintering temperatures. The grains are more in contact and compressed with neighboring grains for more compaction. The higher sintering temperature leads to coalescing of larger grains into smaller grains. The compactness of the samples was displayed in Fig. 5(c). The compactness increases as a function of sintering temperature. The compactness of SPS 900, 1050 is most likely similar to each other based on the error bars. The compactness of SPS 1100, 1150 is in the range of 99.23 and 99.82 %. Fig. 5 (d) illustrates the hardness profiles of the commercial NiTi sample and SPS samples, at fixed load by Vickers hardness testing. Hardness measurements were taken from both the cross-section and surface of the SPS samples and the NiTi shape memory alloy (SMA). It was observed that as the sintering temperature increased, the hardness in both the surface and cross-sectional areas also increased. The average error was calculated from ten indentation points along both the surface and cross-sections. The surface hardness is higher than the cross-section hardness of the SPS sintered samples. This may be due to surface hardness measured on the surface of the samples. However, cross-section hardness is measured through the cross-section of samples, which contains porosity along the thickness. The sample SPS 1050 has lower hardness compared to other samples that arise from the quantity of porosity within the sample.

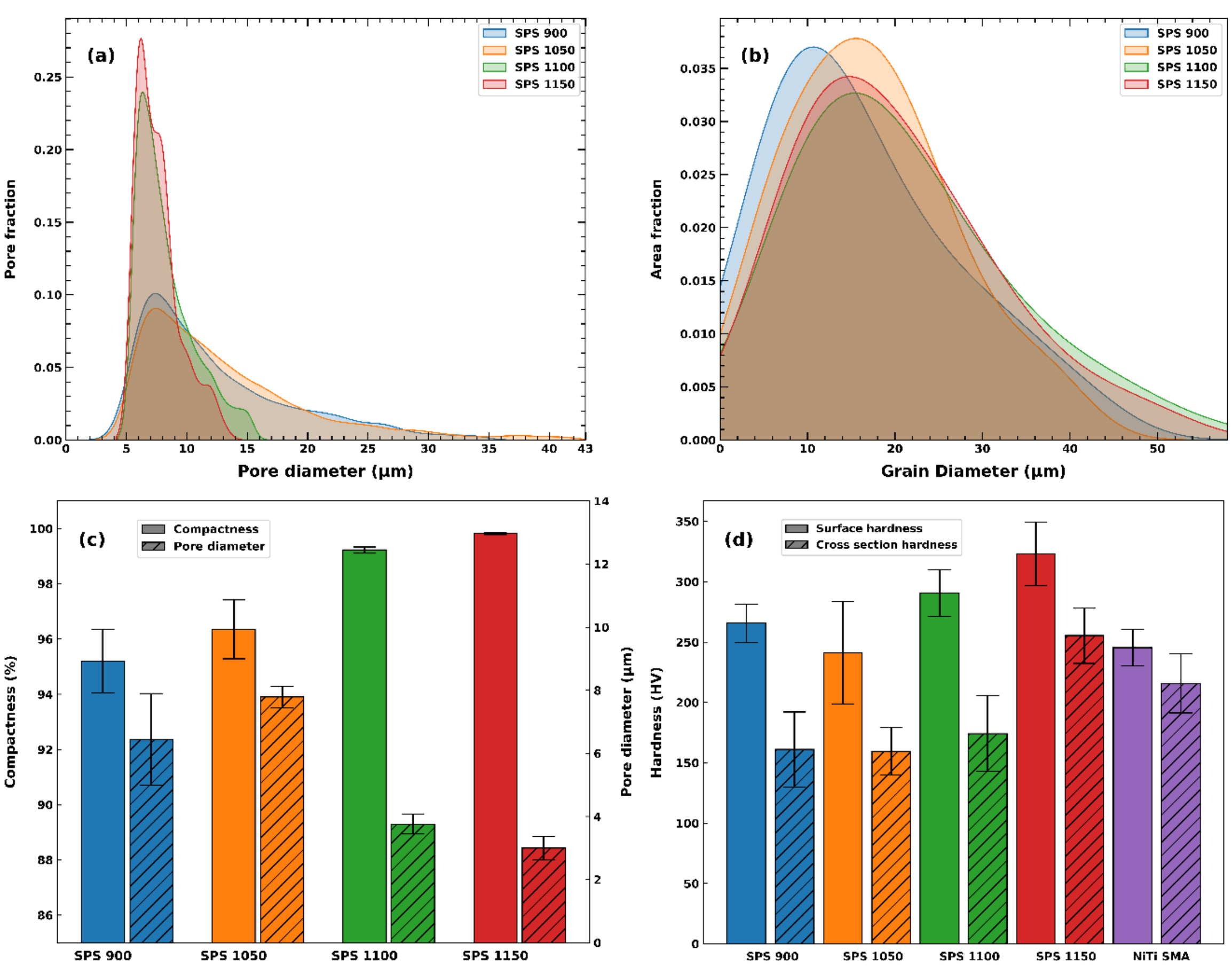

Fig. 5. Quantitative evaluation of a) Porosity distribution for all the fabricated samples, b) Grain size distribution of the samples as a function of area fraction, c) Compactness versus pore diameter for fabricated samples , d) Hardness of the fabricated samples at surface and cross-section areas and compare with commercial NiTi SMA sheet.

### 3.5. Texture orientation of the sample in TD and RD

The most compact sample SPS 1150 is chosen for texture anisotropy orientation in both RD and TD directions. Polycrystalline grains are formed within coarse particles that may be generated during compaction under load and sintering temperature. The multiple grains are more refined in the TD sample compared to the RD sample, which may be due to the applied load along the normal axis of the applied load. Fig. 6 (b, e) displays the graphs of disorientation versus the number of fractions for polycrystalline grains in the TD and RD samples. The disorientation profile in TD samples follows the Gaussian trend, and the grain orientation in the RD sample shows non-uniformity. This may lead to uniform distribution of load and temperature at the center position in the TD sample, which may align along the grain alignment direction, so the misorientation follows the trend. However, the samples in the RD position are positioned perpendicular to the applied load, as a result, misorientation of grains is more irregular and follows a non-uniformity pattern in the sample. The grain size distribution of the selected area shows two zones of grain distribution in coarse and fine regions in the TD sample, which are clearly in the range of 5-25 μm and 30-45 μm. However, the selected position in the RD sample shows multiple grain diameters, finer, medium and coarser ranges.

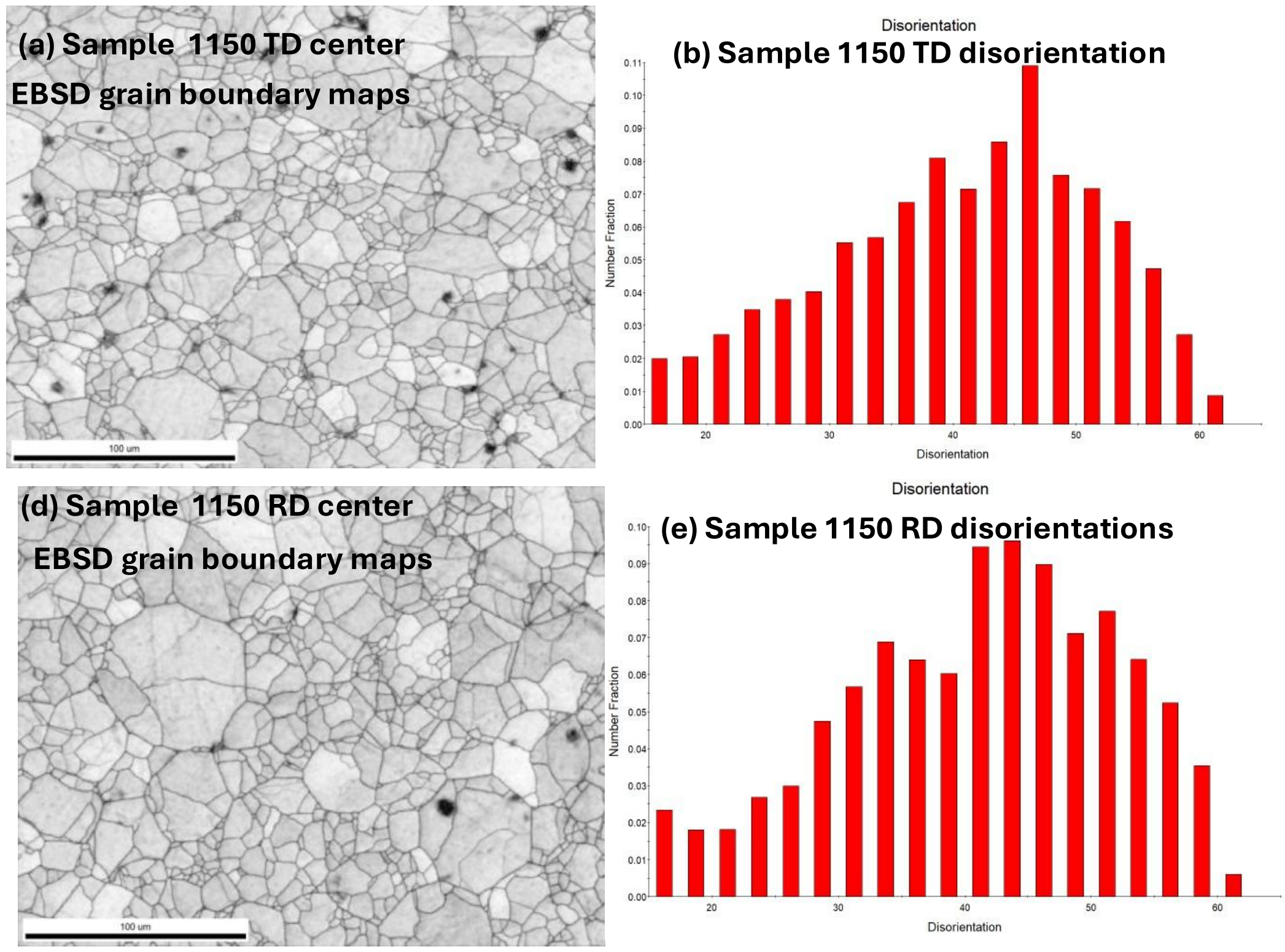

Fig. 6. (a) EBSD grain boundary map, (b) Disorientation graphs, at center position for RD and TD of the sample SPS 1150.

The corresponding inverse pole figure is displayed in Fig. 7 (a-b) for both the RD and TD samples. The harmonic texture of RD is greater than that of TD, which corresponds to the orientation of the [010] direction. However, the texture difference in the TD and RD samples is not a remarkable difference. The minimal difference is also observed in the mechanical response. The unrecovered strain in TD and RD responses of sample 1150 has slightly changed from 0.22 to 0.15 %.

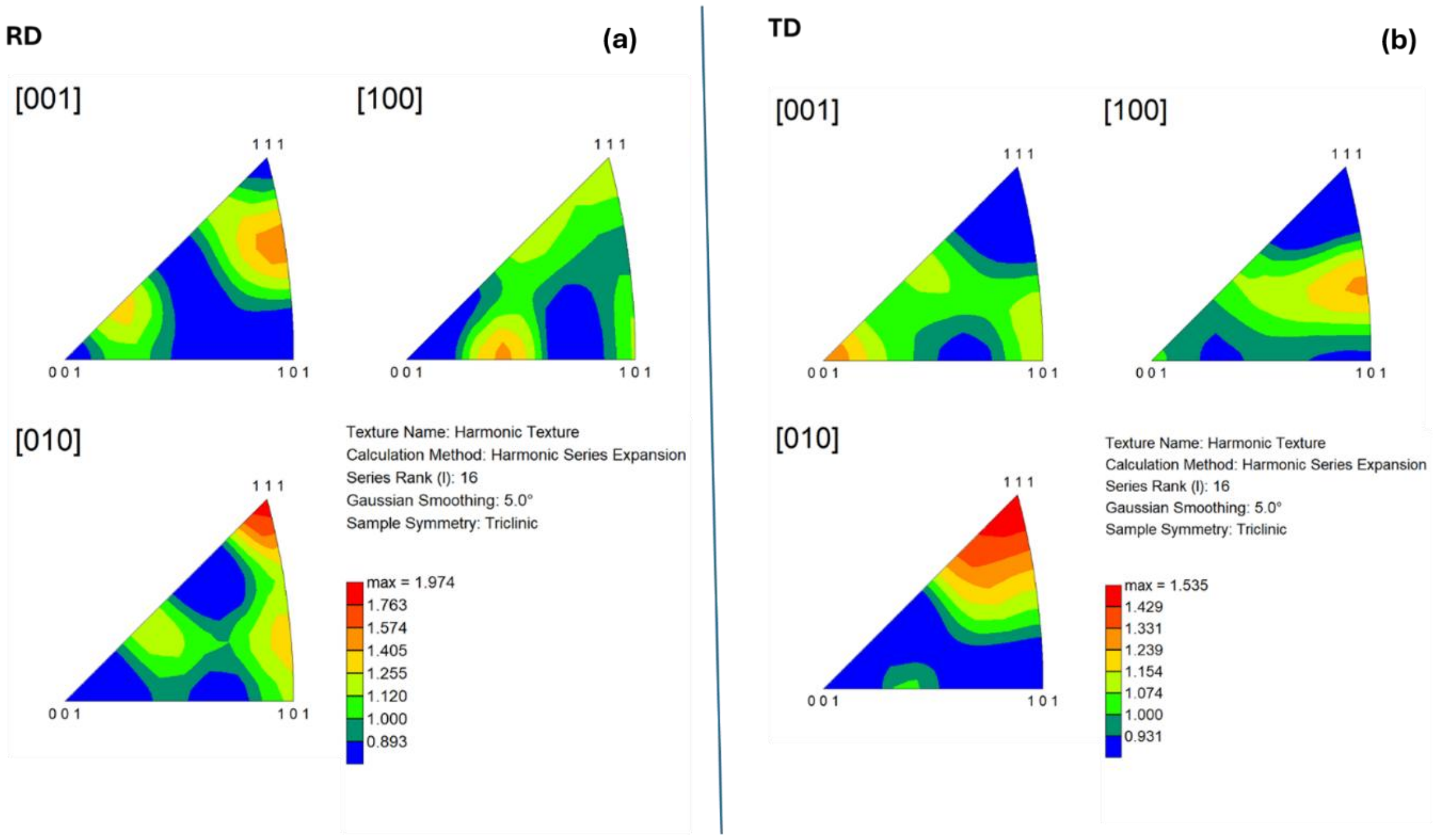

Fig. 7. Inverse Pole Figure (IPF) of the center of a ) RD and b) TD of SPS 1150 along the direction of [001], [100], [010].

The crystallographic orientation of the elements was investigated in RD, showing a higher value of texture compared to TD.

The mechanical behavior of the sample in TD direction shows comparatively better results than RD sample due to the alignment of grains along the normal axis and the texture is at lower values. However, RD sample shows a discontinuous, non-uniform distribution of grains, and the texture value is higher, which leads to a lower mechanical response. The distribution of larger and smaller particles exhibits two different phenomena during the sintering process. The larger particles undergo plastic deformation under mechanical load during the sintering process; however, smaller particles undergo a heating effect with mechanical movement for more compaction. In the case of larger particles, during the sintering process, many grains are formed within the particles. The transformation of spherical particles to hexagonal grain sizes occurs during the sintering process [15]. This plastic deformation leads to the formation of grains, which is observed in Fig. 7 (a, d).  The compaction of particles leads to densification and an increase in hardness [16]. The grains are uniformly oriented in the TD center of sample SPS 1150. At RD, the larger and smaller grains are distributed uniformly in sample SPS 1150. The harmonic texture at TD is lower than RD sample, which leads to better mechanical performance due to uniform orientation [17-18].

### 3.6. Transformation temperature of various phases by thermal characterization

Fig. 8 (a-d) represents the transformation temperature of the NiTi powder, samples prepared by the SPS technique, and annealed powder and samples. On cooling cycles, the R-phase starting temperature ($R_s$) is −29.34 °C and a martensite finish temperature ($M_f$) of −71.86 °C and on the heating cycle the R-phase and austenite finish temperature ($A_f$) are observed at 43.8 °C for NiTi powder. However, samples fabricated using SPS show the merged peaks of R phase and martensite phase during the cooling cycle and R phase and austenite during the heating cycle. During the cooling cycle R phase starts at 13.91 and the martensite phase finishes at 54.03 and on the heating cycle R phase starts at 61.52 and the austenite phase finishes at 49.14 °C for sample SPS 1100. The sample is a mixture of R phase and austenite at room temperature. However, sample 1150 shows R phase at 39.50 and martensite phase at −86.43 during the cooling cycle and R phase starts at 58.44 and austenite finishes at −1.71°C during the heating cycle. Table 3 represents the various phase transformations of R phase, austenite and martensite for NiTi powder and samples as prepared and after annealed NiTi powder and the annealed samples.

Table 3. Phase transformation of R phase, austenite and martensite for NiTi powder and fabricated samples and the corresponding annealed NiTi powder and annealed SPS are presented.

| Sample | Cooling Transformation temperature ( ºC) | | | | Heating Transformation temperature ( ºC) | |
|---|---|---|---|---|---|---|
| | $R_s$ | $R_f$ | $M_s$ | $M_f$ | $A_s$ | $A_f$ |
| NiTi powder | | | 9.06 | −56.31 | - | 43.26 |
| Sample 900 | | | −38.8 | −74.3 | - | 16.8 |
| Sample 1050 | | | −33 | −72 | - | 10 |
| Sample 1100 | | | −39.61 | −93.18 | - | 0 |
| Sample 1150 | | | −30.9 | −84.5 | - | 0 |
| Annealed NiTi powder | 55.25 | - | - | −15.08 | 20.56 | 69.3 |

| | | | | |
|---|---|---|---|---|
| Sample 900 | 43.3 | –48.5 | –13.86 | 58.3 |
| Sample 1050 | 44.2 | –59.8 | –12.9 | 60.2 |
| Sample 1100 | 35.9 | –59.6 | –27.04 | 57.3 |
| Sample 1150 | 39.86 | –59.01 | –14.83 | 51.15 |

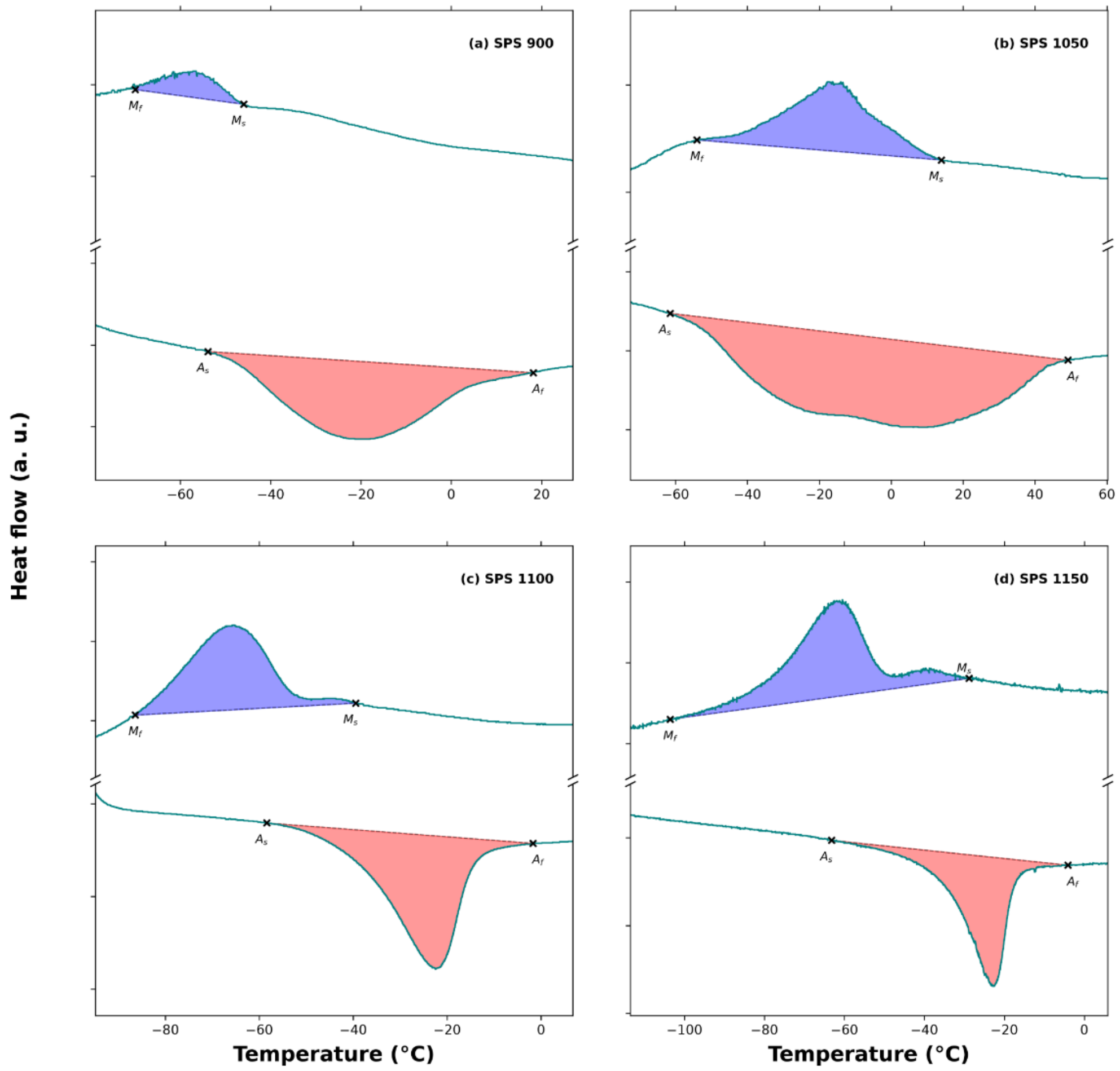

Fig. 8. Phase transformation temperature by DSC graph left column samples prepared at different temperatures at 900, 1050, 1100, 1150 °C and NiTi powder.

### 3.6. Mechanical behavior of samples in TD and RD

The rectangular samples were prepared from bulk samples after electric discharge machining. Fig. 9 (a-d) displays the tensile response of the fabricated samples at room temperature. The samples SPS 900, SPS 1050 and SPS 1100 show the failure in the first cycle as a result of stress-strain behavior. The sample SPS 900 shows failure at the initial stage of strain 0.25 % with an inset image showing the fractured sample that may be developed due to the porous nature of the sample ( Fig. 9 a ). The sample SPS 1050 shows failure at 0.5 % strain with stress of 200 MPa, which also corresponds to the quantity of pores present in the sample. However, sample SPS 1100 although it reflects improved mechanical property of tensile stress of 470 MPa with a strain of 0.8 %, still the sample breaks down at the neck point. The most compact sample SPS 1150 shows super elastic response in the first cycle with fully recovered strain , however the sample does not survive in the second cycle. The failure of the sample occurs at the rounding-flat interface, where the stress is expected to be higher. It might be related to "retained" dislocations and residual stresses. The superelasticity has been confirmed from the mechanical test in the most compact sample.

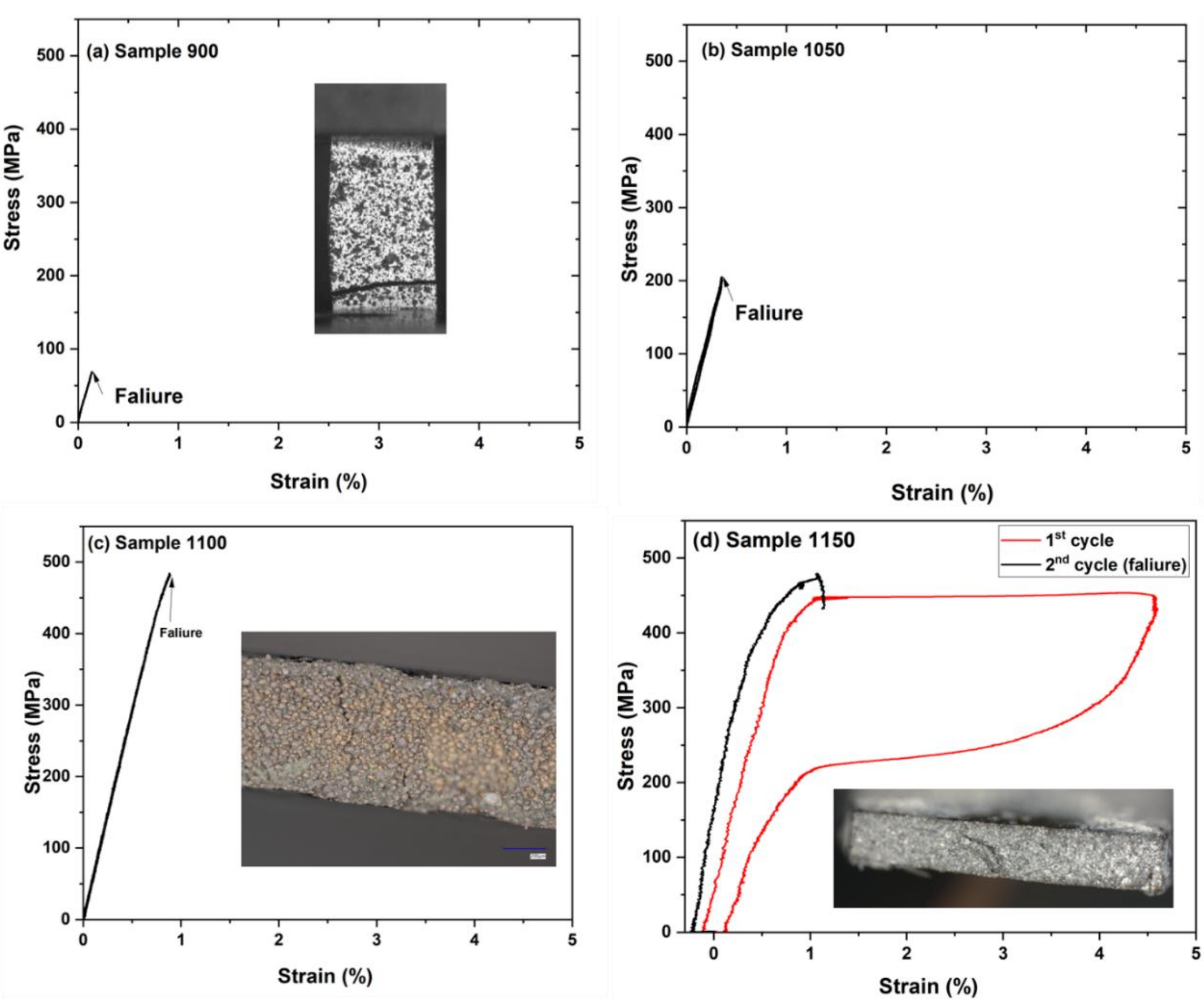

Fig. 9. Tensile test of the as-prepared samples represents a) SPS 900 failure at 0.25 % strain, b) SPS 1050 shows failure at 0.5 % strain, c) SPS 1100 shows failure at 0.8 % strain, d) SPS 1150 shows SE at first cycle and failure at second cycle.

Dynamic mechanical analysis (DMA) was performed on the most compact samples during thermal cycling from -140 to 180 °C to evaluate the temperature dependence of the storage and loss moduli, tan delta, and strain. Figure 10 (a,b) shows the corresponding results for the SPS 1100 and SPS 1150 samples.

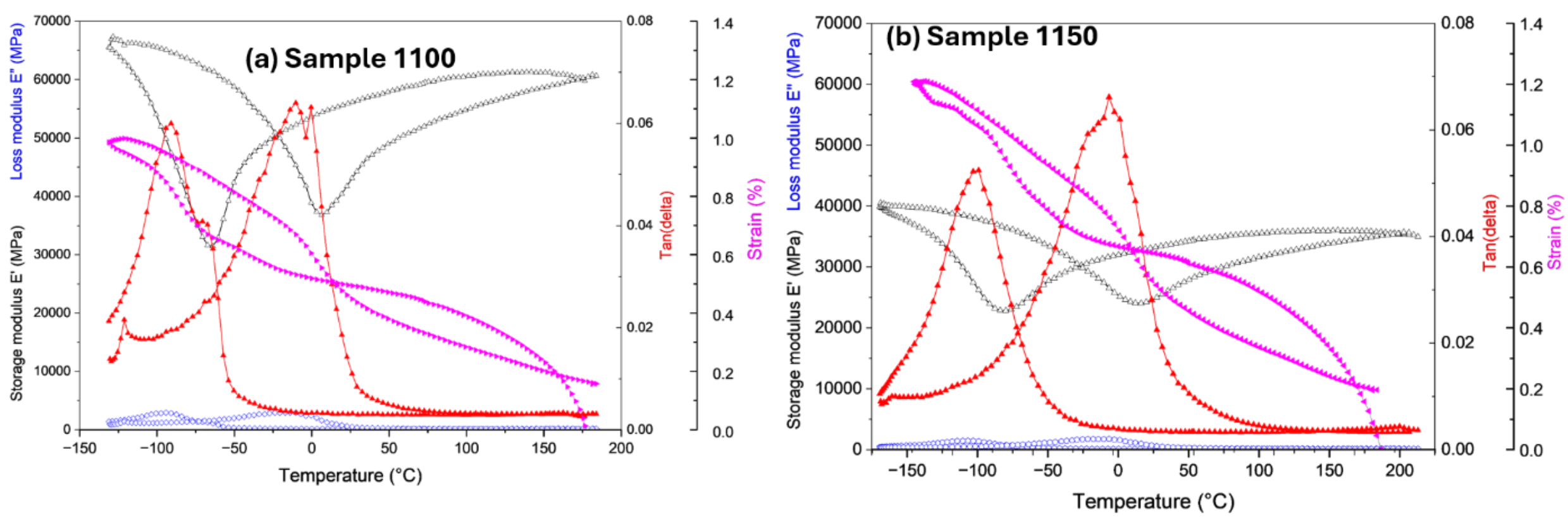

Fig. 10. Storage modulus, loss modulus, tan delta, and strain ~~(%)~~ as a function of temperature for the SPS 1100 (a) and SPS 1150 (b) samples

The storage modulus of SPS 1100 sample was approximately 65 GPa, whereas SPS 1150 sample exhibited a lower storage modulus of approximately 40 GPa. The tan delta curves displayed two characteristic peaks corresponding to the forward and reverse martensitic transformations, in good agreement with the transformation temperatures determined by DSC analysis. The transformation temperatures evaluated from DMA are also consistent with those identified by thermomechanical analysis (TMA). During cooling under constant tensile stress, the strain gradually increased to approximately 0.8 % as a result of the temperature-induced phase transformation. The strain was fully recovered upon subsequent heating, demonstrating a completely reversible thermomechanical response in both investigated samples.

Fig. 11 represents the deformation behavior of the as-fabricated sample SPS 1100 and 1150 and their corresponding annealed samples under fixed load as the function of thermal cycles.  The samples under three-point bending test show deformation during cooling ( by converting the martensite phase B19′) and return to the initial position by converting to austenite phase ( B2) during heating. The change in displacement shows the hysteresis behavior of the sample in Fig. 11. However, the hysteresis is much wider in the annealed samples ( SPS 1100 and SPS 1150 were annealed to 500 °C for 1 h). These samples undergo bending and show the transformation of B2 phase to R phase and then martensite phase B19‘ during cooling and returning from B19′ to B2 austenite phase during heating. In the annealed samples R phase is clearly observed; that's the reason the hysteresis is much wider, the sample transforms through R phase from B2 to B19′.

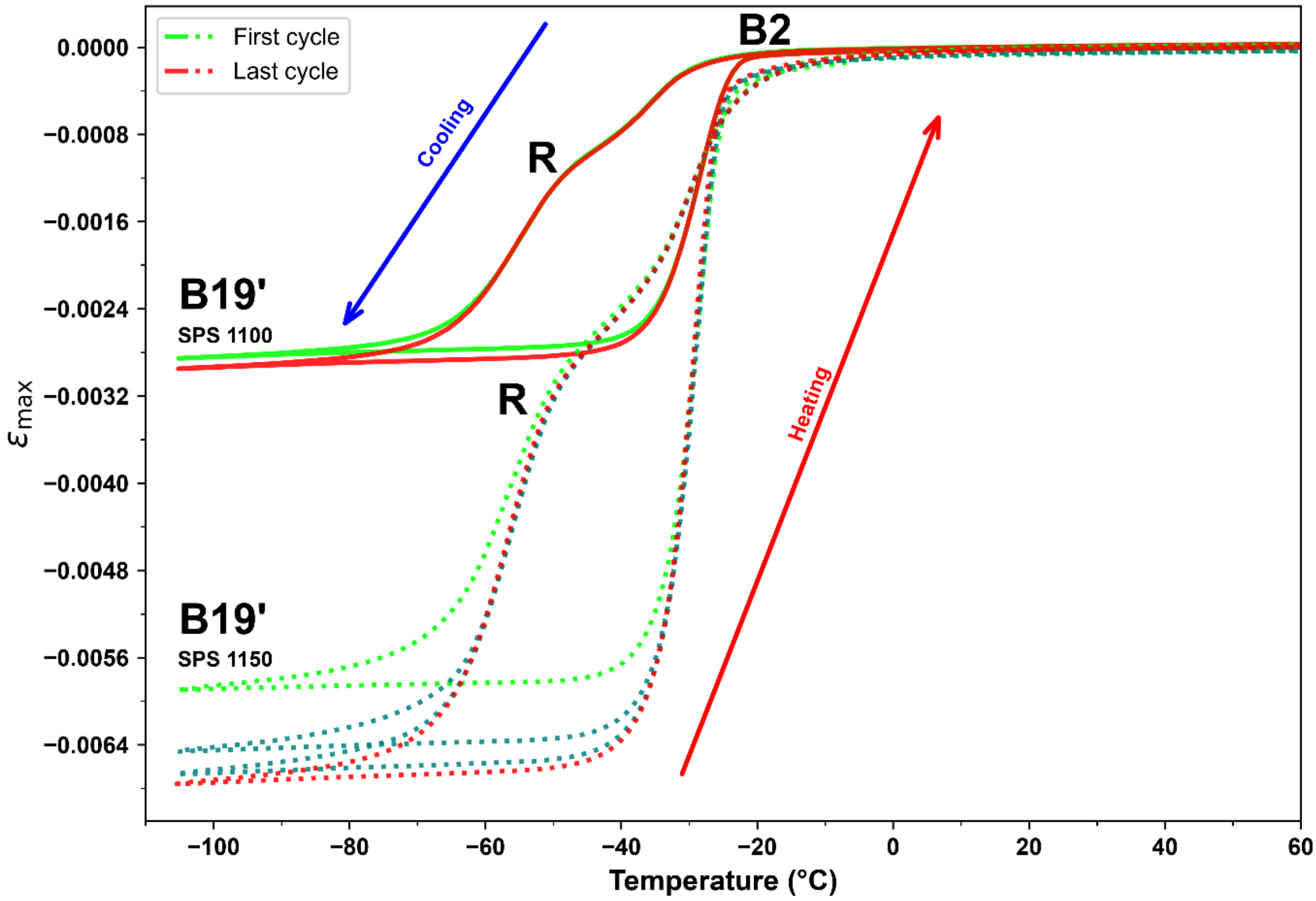

Fig. 11. Samples under 3-point bending under the fixed load of 200 mN: a) fabricated sample SPS 1100 and annealed sample under bending shows 5 cycles.

## 4. Discussion

The sintering mechanism of Shape memory alloy NiTi is primarily controlled by the densification process and anti-densification behavior of the material. During sintering, densification and anti-densification proceed concurrently, closely correlating with microstructural evolution that is predominantly influenced by sintering parameters. Anti-densification, or the persistence of porosity, in NiTi shape memory alloys (SMA) during Spark Plasma Sintering (SPS) is heavily influenced by sintering temperature, which governs atomic diffusion and phase formation. While SPS enables rapid densification (e.g., ~2.5% porosity at 900 °C), lower temperatures can lead to higher porosity due to insufficient atomic diffusion (Fig. 7), while excessive temperatures can trigger undesirable reactions and phase changes that introduce new pores ( Fig. 11). Influence of temperature on porosity and densification.

- Low Temperature Sintering (<900°C): Results in higher porosity due to incomplete neck formation between powder particles.
- Intermediate Temperature Sintering (900–1050°C): Produces dense NiTi alloys (around 97.5% density) by promoting plastic deformation and diffusion within the powder particles, aided by Joule heating.
- High Temperature Sintering (>1050°C): While generally reducing overall porosity, excessive heat can cause the formation of undesirable brittle phases ( Fig. 4, 5).

Anti-Densification (Swelling) Factors: Phase Transformation Swelling**:** Reactive sintering forms intermediate phases before achieving the final phase, which can cause volumetric changes; Gas Entrapment**:** If sintering occurs in an improper atmosphere, gases can become trapped in isolated pores, preventing complete compaction.

- Optimal Sintering Strategy**:** To achieve high density and shape memory properties, sintering is often performed at high temperatures (e.g., 900–1150°C), sometimes using spark plasma sintering (SPS) to override anti-densification forces, achieving densification up to 99.7%.

- Final Stage**:** High-temperature homogenization. If too high or too long, grain growth occurs, causing pores to become trapped inside grains, reducing further densification potential ( Fig. 5, 11).

The particle sintering model displayed in Fig. 12 shows the mechanism of sintering during compaction by the SPS method.

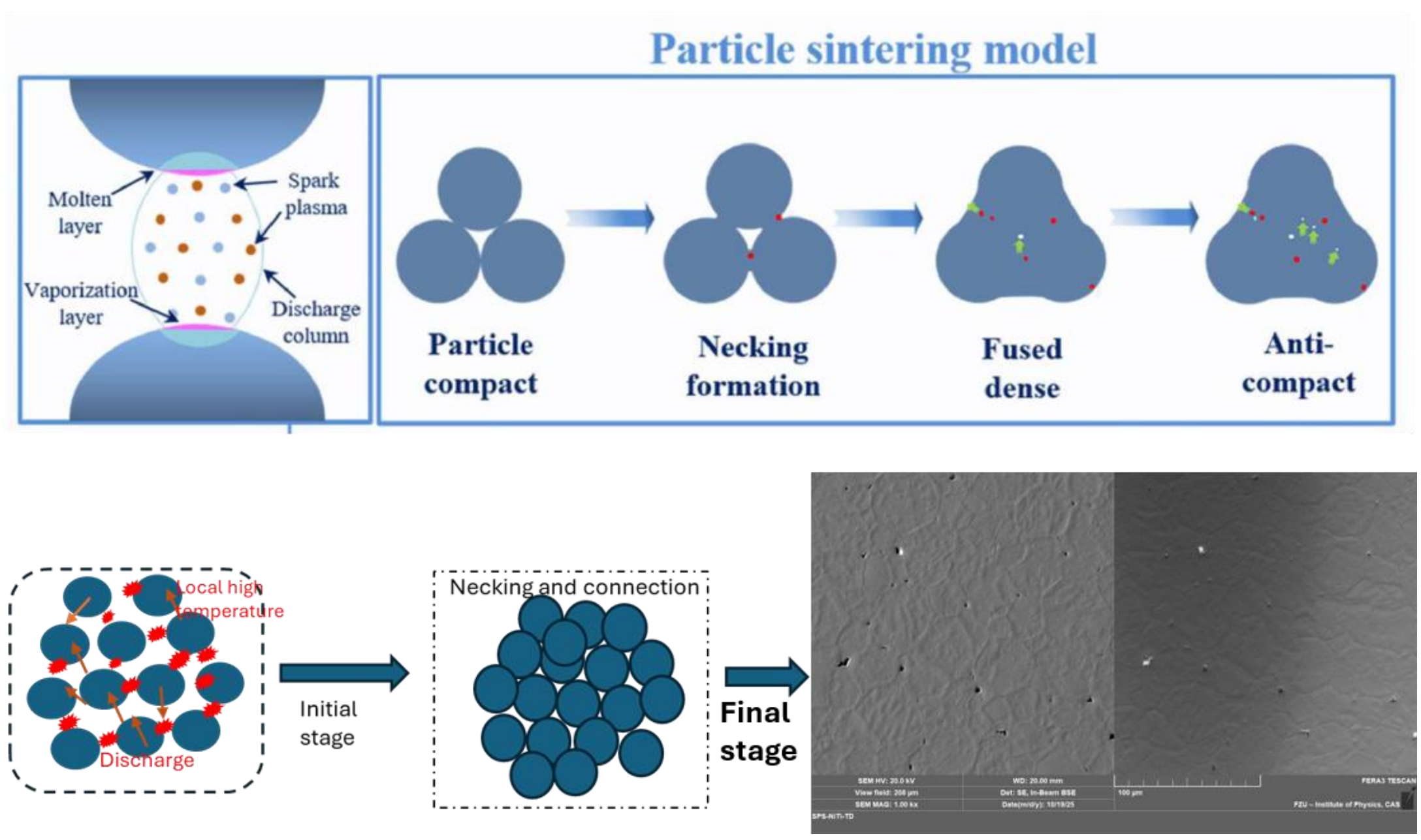

Fig. 12. Schematic diagram for the sintering model of particle model in spark plasma sintering method.

The primary densification mechanisms in SPS of NiTi are particle rearrangement, plastic deformation of particles, and surface diffusion, which are enhanced by the combined effect of high uniaxial pressure and Joule heating.

## 5. Conclusion

The mechanical analysis of fabricated SPS samples was correlated with compaction, microstructure, texture evolution, phase transformation of the fabricated SPS samples. The key findings are as follows.

1. The mechanical response shows SE behavior and bending shows a shape memory effect from the hysteresis response of the sample.
2. The storage modulus, loss modulus, and tan delta were observed in the sample and match well with DSC findings and there is 0.8 % strain recovery in both samples.
3. The mechanical response of the most compact sample shows SE effect that corresponds to less porosity. The densification and hardness increase with a decrease in porosity for the fabricated SPS sample with an increase in sintering temperature from 900 to 1150 °C at a fixed load of 50 MPa.
4. The texture evolution investigates the random distribution of grains in the anisotropy direction.
5. The as-fabricated sample and annealed samples both show hysteresis effect with recovery on heating reflecting the shape memory behavior.

**Acknowledgment**

Part of the work was performed during an open science program internship by the student Esme Erika Kalovcova for laser cutting the sample for texture analysis.

**Declaration of Competing Interest**

The authors declare that they have no known competing financial interests or personal relationships that could have appeared to influence the work reported in this paper.

**Data Availability**

Dataset of Mechanical-microstructural correlation on SPS-fabricated NiTi alloy is available from Zenodo. DOI. 10.5281/zenodo.21869098

**Funding**

MEYS of the Czech Republic is acknowledged for the support of infrastructure projects, CNL (CzechNanoLab LM2023051) and FerrMion (CZ.02.01.01/00/22_008/0004591).